\documentclass[aps,pre,twocolumn,showpacs,amssymb]{revtex4}

\usepackage{graphicx}
\usepackage{dcolumn}
\usepackage{bm}
\def\gs{\gtrsim}
\def\ls{\lesssim}
\def\be{\begin{equation}}
\def\en{\end{equation}}    
\def\gs{\gtrsim}
\def\ls{\lesssim}

\def\p{\partial}
\def\bea{\begin{eqnarray}}
\def\ena{\end{eqnarray}}

\begin{document}
\preprint{APS}
\title{Bubble formation in water 
 with addition of a hydrophobic solute }

\author{Ryuichi Okamoto$^1$}
\author{Akira Onuki$^2$}
\affiliation{
$^1$Department of Chemistry, Tokyo Metropolitan University, 
Hachioji, Tokyo 192-0397, Japan\\
$^2$Department of Physics, Kyoto University, Kyoto 606-8502, Japan}

\date{\today}

\begin{abstract}  
We show that phase separation   can occur  
 in a one-component liquid   outside its  coexistence curve (CX)  
with addition of a  small amount of a   solute. 
The solute concentration at the transition  
decreases with increasing the difference of the 
solvation chemical potential between  liquid and gas. 
As a typical bubble-forming solute, we consider   O$_2$ in ambient 
liquid water, which exhibits  mild  hydrophobicity  
 and its critical temperature is  
lower than that of water. Such a solute can be  expelled from the liquid 
 to form   gaseous domains  while   the  surrounding 
 liquid pressure  is  higher than  
 the saturated vapor pressure 
$p_{\rm cx}$. This   solute-induced bubble formation is 
 a  first-order transition in bulk and 
on a  partially  dried wall, 
while a gas film grows  continuously    
 on a  completely  dried wall.  
 We set up a bubble free energy $\Delta G$ 
for bulk and surface bubbles with a small volume fraction $\phi$. 
It becomes a function  of the bubble radius $R$ 
under the Laplace pressure balance. 
Then, for sufficiently large solute densities above a threshold, 
$\Delta G$  exhibits a local maximum at a critical radius  
and a minimum  at an equilibrium radius. 
 We also examine    solute-induced  nucleation  taking place outside CX, 
where bubbles  larger than the critical radius grow 
until attainment of equilibrium.
\end{abstract}
\maketitle

\section{Introduction}

Recently, much attention has been 
paid to the formation of 
small bubbles, sometimes called nanobubbles,  in water 
 \cite{review1,bridging1,review2}. 
They have been observed  with a dissolved gas 
on hydrophobic surfaces 
 \cite{review2,review1,bridging1,b0,b00,b1,ex10,ex11,ex12,ex13,Ducker,type} 
 and in bulk \cite{Jin,ex1,ex2,ex3,ex4,Bunkin} 
  in  ambient conditions (around $300$ K and 1 atm), 
where the pressure in the bulk liquid region is larger 
than the saturated vapor pressure $p_{\rm cx}$ or  outside 
  the coexisting curve (CX).   
Their typical radius $R$ is   of order $10-100$ nm  
and their life time is   very long.
The interior pressure is 
given by $2\sigma/R \sim$30 atm for a bubble with $R=50$ nm 
from the Laplace law, where $\sigma$ 
is the surface tension equal to  $72$ erg$/$cm$^2$.    
Strong attractive forces 
have also been measured 
 between hydrophobic walls in water  
due to bubble bridging \cite{bridging1,b0,b00,b1,ex1,ex10,ex12}. 
These effects are   important in various  applications, 
but the underlying physics  has  not yet been well 
understood.

In this paper, we ascribe  the origin of  bubble formation 
 to  a  hydrophobic  interaction   between water and solute 
\cite{Ben,Guillot,Paul,Pratt,Chandler,Garde}.  
In our theory,  the solute-induced phase separation  
generally occurs  in equilibrium  
when the solvent is in a liquid state  outside CX    and 
the solute-solvent interaction is  repulsive. 
Most crucial in our theory 
is the solvation chemical potential of the solute  $\mu_{\rm s}(n,T)$ 
depending   on the solvent density $n$ and the temperature $T$.
With   increasing  such a repulsion,  
its difference $\Delta\mu_{\rm s}$  between the liquid and gas phases 
 can be   considerably 
larger than the thermal energy $k_BT$ (per solute particle). 
In this situation,  the solute molecules  are 
   repelled from the liquid   
to form  domains of a new phase 
(in gas, liquid, or solid).  
 Supposing   bubbles  
with a small volume fraction $\phi$, we   set up a  free energy 
$\Delta G$  accounting for considerably large $\Delta\mu_{\rm s}/k_BT$.  
Then, its minimization with respect to $\phi$ 
and the interior solute density $n'_{\rm I}$ 
yields the equilibrium conditions of 
 bubbles in liquid (those of chemical equilibrium and pressure balance). 

As a bubble-forming solute in water,  we treat O$_2$,  
which is mildly hydrophobic with 
$\Delta\mu_{\rm s}/k_BT \cong 3.44$ on CX at $T=300$ K. 
Furthermore,  the critical 
 temperature and  pressure  
of   water and O$_2$ are given by   
$(647.3$ K,  22.12 MPa) and ($154.6$ K,  $ 5.043$ MPa),  
respectively. Notice that  the  critical temperature of water  
 is much higher  than that of 
O$_2$ (and than those of 
 N$_2$, H$_2$, and Ar etc) due to the hydrogen 
bonding in water. As a result, no gas-liquid phase transition 
takes place within  
 bubbles   composed mostly of  O$_2$ 
 in liquid water in ambient conditions. 
In contrast,    strongly  hydrophobic solutes    
 usually  form   solid aggregates  in liquid  water 
except for very small solute densities 
\cite{Ben,Guillot,Paul,Pratt,Chandler,Garde}.

In our theory,   solute-induced   bubbles can appear 
outside CX  only when the solute density 
exceeds a threshold density, where   
the  threshold  tends  to zero as the 
liquid pressure approaches $p_{\rm cx}$. 
 In particular, above the threshold density,  
 a surface bubble (a gas film) appears on a hydrophobic wall 
in the temperature range $T<T_{\rm D}$ ($T>T_{\rm D}$). 
As is well known, this is possible only on CX 
without solute (in one-component fluids).  
Here,  $T_{\rm D}$  is the drying temperature \cite{Cahn,Bonnreview} 
determined by the solvent-wall interaction, so it 
 is insensitive to a small amount of solute with mild 
$\Delta\mu_{\rm s}$. 
With a solute below the threshold density, we predict only 
 a microsopically depleted layer 
  outside CX  (as in one-component fluids). 
Indeed,    some groups \cite{Doshi,G1,G2} detected  
only microscopic depletion layers 
  on a hydrophobic wall with a dissolved gas, 
while  other groups observed  
surface bubbles 
 \cite{review2,review1,bridging1,b0,b00,b1,ex10,ex11,ex12,ex13,Ducker,type}.

On the other hand, to prepare stable  bulk bubbles, 
macroscopic  gas bubbles composed of O$_2$  etc. 
have been  fragmented  by stirring 
in liquid water \cite{Bunkin,ex2,ex3,ex4}.  
In   such measurements, Ohgaki {\rm et al.}\cite{ex2}  
realized bubbles with $R\sim 50$ nm 
 in  quasi-steady states, where 
the bubble  density was   $n_{\rm b}\sim 
19~\mu$m$^{-3}$ and  the bubble volume fraction 
was  $\phi\sim 0.01$.  
 We shall see that the nucleation  
barrier of creating  solute-induced bulk bubbles 
in quiescent states is    
too  high for nucleation experiments 
for  a gas such as O$_2$ except for very 
high liquid pressures.

As a similar bulk phenomenon, long-lived heterogeneities 
have also been observed  in one-phase states of 
aqueous mixtures  with  addition of a  salt or 
a hydrophobic solute \cite{S1,Curr,Okamoto}. 
Dynamic light scattering 
experiments indicated that 
their typical size is of order $100$ nm.
Theoretically, such a phase separation  can occur 
if the solute-solvent interaction 
is highly  preferential 
 between the two solvent components  \cite{Curr,Okamoto}.

This paper is organized as follows. In Sec.II, we 
will present a thermodynamic theory 
 of bubbles induced by a small amount of solute, where 
the liquid  pressure  and the total solvent and solute  numbers 
are fixed. In Sec.III, we will set up a bubble free energy $\Delta G$. 
In Sec.IV, we will examine  solute-induced nucleation.
In addition, in Sec.IIIC and Appendix A, we will briefly 
 examine bubble formation 
at fixed chemical potentials and 
at fixed cell volume.

\section{Equilibrium  bubbles with hydrophobic solute }

We consider  a one-component solvent, called water,  
in  a liquid  state outside the coexistence curve (CX). 
We then add  a  small  amount of a neutral, hydrophobic solute  
  (impurities).  
The total solvent and solute 
numbers are fixed at $N= V{\bar n}$ and $N_{\rm I}= 
V{\bar n}_{\rm I}$, 
respectively, with   $\bar n$ and ${\bar n}_{\rm I}$ being  
the  initial water and solute  densities.   
Here,  ${\bar n}$ is larger than     the liquid density
  $  n_{\rm cx}^\ell$ on CX before  bubble formation.  
 We  keep the pressure in the liquid region 
 at the initial value 
$\bar p$ larger than     the saturated vapor (coexistence)  
pressure  $p_{\rm cx}$ 
by attaching a pressure valve to the cell, as illustrated in Fig.1. 
We do not assume the presence of  surfactants and ions 
(see remarks in Sec.IVA).

\subsection{Solvation chemical potential 
and Henry's law}

We  assume  that the  molecular 
volume  of solute  $ v_{\rm I}$  
is of the same order as that of solvent 
   $v_{\rm w}$, since 
 large hydrophobic impurities tend  to form solid precipitates 
\cite{Paul,Chandler,Garde}.  
We  then consider   the Helmholtz free energy  density $f$ depending 
on  the water density $n$ and the solute density $n_{\rm I}$ 
in the dilute limit of solute. 
In this paper, we neglect  the solute-solute 
interaction to obtain 
\be
f(n,n_{\rm I}) = 
f_{\rm w} (n) + k_B Tn_{\rm I}[ \ln (n_{\rm I} v_{\rm I}) -1 +\nu_{\rm s}(n)]  , 
\en 
where $f_{\rm w}(n)$ is the  Helmholtz free energy density of  pure water  
and $\nu_{\rm s}(n)$  is related to the solvation  chemical potential 
$\mu_{\rm s}(n)$ in the limit of small $n_{\rm I}$ by 
\be 
\nu_{\rm s}(n)= \mu_{\rm s}(n)/k_BT.
\en  
Hereafter, 
the $T$-dependence of the physical  quantities 
will not be  written explicitly.  

Note that the combination  $\ln [v_{\rm I}/\lambda_{\rm I}^3]
+ \nu_{\rm s}$ can be determined unambiguously in 
thermodynamics in the limit $n_{\rm I}\to 0$, where 
$\lambda_{\rm I}$ is the thermal de Broglie  
length ($\propto T^{-1/2}$).  Thus, 
  $v_{\rm I}$ may be chosen  to be  independent of $n$ 
without loss of generality.  
It is known that the entropic contribution to  
 $\nu_{\rm s}$ is crucial 
 for  nonpolar impurities  in liquid water 
\cite{Pratt,Paul,Chandler}. 

From  eq.(1) we calculate the  chemical potential of water $\mu$ 
and that of solute $\mu_{\rm I}$ as  
\bea 
&&
\mu=
\p f_{\rm w}/\p n+ k_B T n_{\rm I} g_s(n),\\
&& \mu_{\rm I}= 
k_B T [\ln (n_{\rm I} v_{\rm I})+ \nu_{\rm s}(n)],
\ena  
where we define 
\be 
g_s(n)= \frac{\p \nu_{\rm s}}{\p  n}= \frac{1}{k_BT} 
 \frac{\p \mu_{\rm I}}{\p n}.
\en 
The pressure $p= n\mu+n_{\rm I}\mu_{\rm I}-f$ is written as 
\be 
p= [n\p f_{\rm w}/\p n-f_{\rm w}]+ k_BT n_{\rm I}[1+ng_s(n)],
\en  
where the first term is 
the  contribution from the solvent and the second term from the solute. 
The   typical size of  $g_s(n)$ is 
of the order of the solute molecular 
volume   $ v_{\rm I}$. In the presence of bubbles, 
 $\mu$ and  $\mu_{\rm I}$ take common values in  gas and  liquid, 
while the pressure in the bubbles is higher than that in the liquid by 
$2\sigma/R$.

First, the homogeneity of  $\mu_{\rm I}$ 
in equilibrium yields  
\be 
n_{\rm I} =n_{\rm I}^0  \exp[  -{ \nu}_s(n)], 
\en 
as a function of $n$ in two-phase states, 
where $n_{\rm I}^0=\exp(\mu_{\rm I}/k_BT)/v_{\rm I}$ is a constant. 
In gas-liquid coexistence, let 
the water and solute  densities be  $n'$ and  
$n_{\rm I}'$ in gas  and be  $\hat n$ and 
 ${\hat n}_{\rm I}$ in liquid, respectively. Then,  eq. (7) gives 
\be  
{ {\hat n}_{\rm I}}/ n_{\rm I}'=  
\exp[ -\Delta{\nu}_s],
\en 
where the $\Delta\nu_{\rm s}= {\nu}_s({\hat n})-{\nu}_s({n'})$. 
This  density ratio  is called  the Ostwald 
 coefficient, which  represents   solubility of a gas 
\cite{Ben,Guillot,Pratt}. It is much smaller  than unity for 
large $\Delta\nu_{\rm s}$. 
Near CX,  we may approximate  $\Delta\nu_{\rm s}$ by its value on CX 
expressed as 
\be 
\Delta\nu_{\rm s}= \nu_{\rm s}(n_{\rm cx}^\ell)-\nu_{\rm s}(n_{\rm cx}^g),
\en 
where  ${n}_{\rm cx}^\ell$ and ${n}_{\rm cx}^g$ are 
the liquid and gas densities on CX of pure water.

It is worth noting  that   $\Delta\nu_{\rm s}$  in eq.(9) is  related to the 
Henry constant $k_{\rm H}$\cite{Sander,Smith}. From partitioning of 
a solute between coexisting gas and liquid, 
it is defined by  
\be  
k_{\rm H}= k_BT n'_{\rm I}/\hat{x}=k_BTn_{\rm cx}^\ell 
\exp({\Delta\nu_{\rm s}}),   
\en  
where   $k_BTn'_{\rm I}$ is the solute partial  pressure in gas 
 and $\hat{x}=\hat{n}_{\rm I}/n_{\rm cx}^\ell$ 
is the solute molar fraction in liquid. 
In  water in the ambient conditions, 
 $\Delta\nu_{\rm s}$ is $3.44$ for  O$_2$, 
   $4.12$ for N$_2$,  and  0.18 
for CO$_2$, where   CO$_2$   is highly  soluble in   liquid water.
Thus, our theory is not applicable to CO$_2$.

However, there are  a variety of 
solutes with stronger  hydrophobicity\cite{Sander}.
For example, $\Delta\nu_{\rm s}=10.2$ 
for pentacosane.  In addition, from 
numerical simulations, 
  a neutral hard-sphere  particle 
deforms the surrounding hydrogen bonding; as a result,  
 $\Delta\nu_{\rm s} \propto a^3$ for $a \ls 1$ nm and 
  $\Delta\nu_{\rm s}  \sim 4\pi\sigma a^2/k_BT$ 
 for $a>  1$ nm  with varying  the particle 
radius $a$ \cite{Paul,Garde,Chandler,Pratt}. 
 This  gives $\Delta\nu_{\rm s}
 \sim 180$ for $a \sim 1$ nm. As  hydrophobic assembly, 
such  strongly  hydrophobic solutes 
 aggregate  in liquid water.  

\begin{figure}
\includegraphics[width=0.9\linewidth]{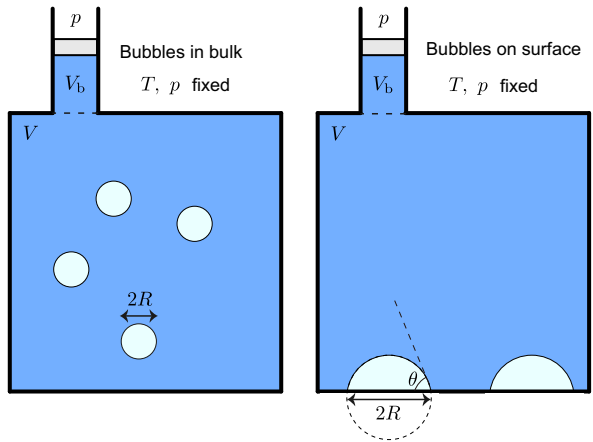}
\caption{(Color online) Illustration of 
experimental setup with a pressure valve 
realizing  a  constant liquid  pressure $\bar p$ 
(outside bubbles) larger than $p_{\rm cx}$, where 
the total solvent and solute numbers are fixed. The cell 
contains bubbles in bulk (left) or those  on a wall (right). 
The volume of the cell is $V$ 
and that of the valve region  is $V_b$, 
where $V_b$ is nearly equal to $\phi V$ for small bubble 
 volume fraction  $\phi$ in the cell. 
}
\end{figure} 

\subsection{Chemical equilibrim and 
pressure balance }
We consider  bubbles in bulk or 
 on a wall  at  a  small volume fraction $\phi$ in the cell. 
For simplicity, we assume no bubble in the valve region in Fig.1. 
If the water density inside the bubbles 
$n'$ is much smaller than $\bar n$, 
the valve volume $V_b$  is given by 
\be 
V_b=V \phi .
\en 
Since the total solvent and solute numbers are fixed, 
the densities in the liquid are given by 
\bea 
&&\hat{n}= \bar{n}- \phi n'\cong {\bar n},\\
&& {\hat n}_{\rm I}={\bar n}_{\rm I}- \phi {n}_{\rm I}' 
\ena 
Hereafter, we  set ${\hat n}= {\bar n}$. 
We also have   $\phi <{\bar n}_{\rm I}/ n_{\rm I}'\ll 1$ 
from   ${\hat{n}}_{\rm I}>0$.  
Thus, the chemical equilibrium condition 
(8) and the  conservation relation (13) give  
\bea 
&&n_{\rm I}'={\bar n}_{\rm I}/[\phi + \exp(-{\Delta\nu}_s)] ,\\
&&{\hat n}_{\rm I}={\bar n}_{\rm I}/[1 + \phi\exp({\Delta\nu}_s)] .
\ena  
The fraction of the solute 
in the bubbles is given by 
\be 
\alpha=\phi n_{\rm I}'/{\bar n}_{\rm I}= \phi/[\phi + \exp(-{\Delta\nu}_s)], 
\en  
which tends to 1 for $\phi \gg  \exp(-{\Delta\nu}_s)$.

We write the value of     the  water chemical potential $\mu$ in eq.(3) 
in the gas  as  $\mu'$ and that in the liquid as $\hat{\mu}$, 
where   $\mu'= \hat{\mu}$ in equilibrium.  
Since ${\bar p}$ is close to $p_{\rm cx}$, 
it is  convenient to 
measure them from the chemical potential $\mu_{\rm cx}$ 
 on CX for pure water. Here,  $n'$ 
is small in the gas and use can be made of  the Gibbs-Duhem relation 
in the liquid. Then, we obtain  
\bea
\mu' &=&    \mu_{\rm cx}+k_BT[\ln (n'/n_{\rm cx}^g)+n_{\rm I}'g_s(n')]  ,\\
\hat{\mu}&=& {\bar \mu}=  \mu_{\rm cx}+ 
({\bar p}- p_{\rm cx})/n_{\rm cx}^\ell, 
\ena 
where  $\hat\mu$ remains equal to the initial value  $\bar\mu$. 
To linear order  in the deviation $n'- n_{\rm cx}^g$  in eq.(17),  
the  chemical equilibrium  condition $\mu'= \hat{\mu}$ yields 
\be 
n'/n_{\rm cx}^g-1 = ({\bar p}- p_{\rm cx})/k_BTn_{\rm cx}^\ell -n_{\rm I}'
g_s(n'),
\en 
In the right hand side of eq.(19),    we may neglect 
the first term for 
$\hat{p}- p_{\rm cx}\ll k_BTn_{\rm cx}^\ell $ and 
the second term for $n_{\rm I}'v_{\rm I} \ll 1$ 
(see the sentence  below eq.(6)). Then, we  find     
\be 
n'=n_{\rm cx}^g=  p_{\rm cx}/k_BT  .  
\en 
  For  one-component  fluids\cite{Katz},
 the pressure  in a bubble has been set 
  equal to  $p_{\rm cx}$  from $n_{\rm cx}^g/n_{\rm cx}^\ell \ll 1$ 
far from the critical point. 
In the  present mixture case, 
the gas pressure is $p'=k_BTn_{\rm I}'+p_{\rm cx}$. 
With the aid of the  Laplace law  $p'
={\bar p}+ 2\sigma/R$, we obtain the pressure balance equation,    
\be 
k_B T n'_{\rm I}= {\bar p}- p_{\rm cx}  +2\sigma/R.  
\en 
Eliminating $n_{\rm I}'$ from eqs.(14) and (21), we may express 
the volume fraction $\phi$ as 
\be 
\phi=k_BT {\bar n}_{\rm I}/({\bar p}-p_{\rm cx}+2\sigma/R)  
-  e ^{-{\Delta\nu}_s}.
\en

From eqs.(20) and (21), we find   $n'_{\rm I}\gg n'$ 
for ${\bar p}- p_{\rm cx}\gg p_{\rm cx}$ 
or for $R\ll  2\sigma/p_{\rm cx}$, where the gas 
consists mostly of the solute. 
For  water at $T=300$ K, 
we have   $p_{\rm cx}=3.6$ kPa and $n_{\rm cx}^g= 
0.86\times 10^{18}/$cm$^3$, where  
$n_{\rm I}' \gg n'$ holds for ${\bar p}\gg 0.0036$ atm 
or for $R<40~\mu$m.   In addition,   at ${\bar p} =1$ atm, 
we have 
$ n'_{\rm I} \cong 2\sigma/k_B T R$   
for $R\ll  1.4$ $\mu$m.


In the limit of $\phi \to 0$ and $R\to \infty$, 
eq.(22) gives   a threshold solute density 
for gas film formation,  
\be
 n_{\rm I}^c =
e^{-\Delta {\nu}_s}({\bar p}-p_{\rm cx})/k_BT ,  
\en  
which  vanishes as ${\bar p}\to p_{\rm cx}$ and 
is small for large   $\Delta {\nu}_s$. 
Here, we introduce the following parameter, 
\be 
\gamma= {\bar n}_{\rm I}/n_{\rm I}^c-1.  
\en 
A  gas film can appear for  $\gamma>0$, 
but bubbles with $R^{-1}>0$ can be stable 
for $\gamma>\gamma_{\rm tr}$ with 
$\gamma_{\rm tr}$ being a positive threshold (see Fig.4). 
 For O$_2$  in water at $T=300$ K, 
we have $ n_{\rm I}^c = 0.78\times 10^{18}({\bar p}-p_{\rm cx})$ cm$^{-3}$ 
with pressures in atm. 
The corresponding oxygen mole fraction is 
$ 2.3\times 10^{-5}({\bar p}-p_{\rm cx})$.


\subsection{Gas film  at fixed pressure}
We consider a gas film on 
a hydrophobic wall, 
where there is no contact between  
the wall  and  the  liquid  phase 
 Setting  $R^{-1}=0$ in eq.(22), 
we obtain  $\phi$ for $\gamma>0$ as  
\bea 
\phi&=& ( {\bar n}_{\rm I}- n^c_{\rm I})k_BT/({\bar p}-p_{\rm cx})\nonumber\\
&=& \gamma \exp(-\Delta {\nu}_s)  .  
\ena 
In this case, we have ${\hat n}_{\rm I}={\bar n}_{\rm I}/(1 +\gamma)$ 
from eq.(15). For $\gamma<0$, water is only   microscopically depleted 
at the  wall, though the depletion layer itself 
can be influenced by the solute\cite{Doshi}.  
In the present isobaric case,  $\phi$ 
increases and even approaches unity as ${\bar p}-p_{\rm cx}\to 0$, 
so we need to require 
${\bar p}-p_{\rm cx}>k_BT ({\bar n}_{\rm I}-n^c_{\rm I})$. 
In contrast, at fixed cell volume, 
 $\phi$ remains small even for 
 ${\bar p}-p_{\rm cx}\le 0$ (see Appendix A).

\subsection{Bubbles with a common  radius at fixed pressure}

We suppose bubbles with a common curvature $R^{-1}$ 
outside CX. 
Using  the relation $k_BT{\bar n}_{\rm I}/({{\bar p}-p_{\rm cx}})= 
(\gamma+1) \exp(-\Delta\nu_{\rm s})$, we rewrite  eq.(22)  as 
\be 
 \phi=\frac{k_BT({\bar n}_{\rm I}- n_{\rm I}^c)}{{\bar p}-p_{\rm cx}}
  \cdot \frac{R-R_c}{R+ \gamma R_c}  ,
\en 
which tends to eq.(25) in the limit $R\to \infty$. Here, we 
introduce the critical radius  $R_c$ defined by 
\be 
R_c= 2\sigma/\gamma ({{\bar p}-p_{\rm cx}}).
\en 
Here, 
$R_c=  1.4/\gamma$ $\mu$m  for ambient water ($300$ K and $1$ atm). 
We need to require $R>R_c$ outside CX  since  $\phi>0$. 
See $R_c$ for O$_2$ in water in  Fig.5(a). 
For bubble nucleation in one-component fluids\cite{Katz,Caupin,Onukibook}, 
 the critical radius 
 is given by 
$R_c= 2\sigma/(p_{\rm cx} - {\bar  p})$ 
 with   $ {\bar p}<p_{\rm cx}$.

We assume $N_{\rm b}$  bubbles in the cell 
neglecting bubble coalescence. 
Then, we express  $\phi$  as 
\be 
\phi= 4\pi R^3 G(\theta )n_{\rm b} /3, 
\en  
where $n_{\rm b}=N_{\rm b}/V$ is 
the bubble density. 
For bulk bubbles we set $G(\theta)=1$. 
For surface bubbles it  is given by  Turnbull's  formula
\cite{hetero,Binder}, 
\be 
G(\theta)=(2-3\cos\theta+ \cos^3\theta)/4,
\en 
where $\theta$ is  the (gas-side) 
contact angle 
in the partial drying condition 
determined by Young's relation,  
\be 
\cos \theta=  (\sigma_{\rm w}^\ell-\sigma_{\rm w}^g)/\sigma.  
\en  
where $\sigma_{\rm w}^\ell$ and $\sigma_{\rm w}^g$ are 
 the free energies per area between the wall and the liquid 
and gas  phases, respectively, and  we assume 
$|\sigma_{\rm w}^\ell-\sigma_{\rm w}^g|<\sigma$. 
Here,  $0\le \theta < \pi/2$  for a hydrophobic wall 
and $\pi/2< \theta \le \pi$  for a hydrophilic  wall. 
As  $\theta\to 0$,  we have 
the complete drying condition  $\sigma_{\rm w}^\ell-\sigma_{\rm w}^g=
\sigma$ at $T=T_{\rm D}$ on CX.  
As  $\theta\to\pi$, the bubbles tend to be 
 detached from the wall, resulting in  
bulk bubbles. Experimentally, $\theta$ for surface bubbles 
has been observed in a range of $10$-$30^\circ$\cite{review2}. 
Note that eqs.(26) and (28) constitute a closed 
set of equations determining the  equilibrium 
radius $R$ for each given 
 ${\bar n}_{\rm I}$,  $\theta$,  and $n_{\rm b}$.

\section{Bubble free energy} 

\subsection{Derivation using grand potential density}

In the geometry in Fig.1 with a pressure valve,  we should 
derive the equilibrium conditions of bubbles  
from  minimization of  the  Gibbs free energy  written as 
\bea 
G&=& F+\sigma S+ (V+V_b){\bar p} \nonumber\\
&=& {\bar G}+\Delta G, 
\ena 
where $F$ is  the Helmholtz free energy (excluding the surface 
contribution here) and $S$ is the total interface  area. 
 For a small volume change 
$V_b \to V_b+d V_b$,   the work exerted  by the fluid 
 to the valve is ${\bar p}dV_b$ at fixed pressure, so we should consider $G$ 
in eq.(31). The second line is the definition of 
the bubble free energy 
$\Delta G$ with  ${\bar G}= 
Vf({\bar n},\bar{n}_{\rm I})+V{\bar p}$ 
being  the initial Gibbs free energy.

In terms of the Helmholtz free energy densities  $f'$ 
in the gas and $\hat{f}$ in the liquid, 
we have $F= V(\phi f'+ \hat{f})$ for the 
total system including the valve region. 
Here, it is convenient to introduce  
 the grand potential density,  
\be 
\omega (n,n_{\rm I}) 
= f -\bar{\mu}n- \bar{\mu}_{\rm I}n_{\rm I} + {\bar p},
\en 
where $\bar{\mu}$ and $\bar{\mu}_{\rm I}$ 
are the initial chemical potentials for water and solute, respectively. 
Using eqs.(12) and (13) we obtain  
\be
{F}/{V} = \phi\omega'+  \hat{\omega} -\phi{\bar p}
+  {\bar f} ,  
\en 
where $\omega'$ is the value of $\omega$ in the gas,  
${\hat\omega}$ 
is that in the liquid, and  $\bar f$ is the initial 
Helmholtz free energy density.  Thus,  eq.(31)  gives    
\be
\Delta G =V[\phi\omega'+  \hat{\omega}] +\sigma S .
\en

We note that $\omega ({n},{ n}_{\rm I})$ 
vanishes in the initial state  
and is  second order  with respect to the deviations 
$n-{\bar n}$ and $n_{\rm I}- {\bar n}_{\rm I}$ (see eqs.(12) and (13)).  
In the following we assume that  the bubbles  have  
a common curvature  $R^{-1}$, where in terms of $G(\theta)$ 
in eq.(29) $S$ is given by \cite{hetero,Binder}, 
\be
 S =4\pi   R^2 G(\theta) n_{\rm b}V.
\en

We next calculate $\omega'$ and $\hat{\omega}$ 
for small ${\bar p}-p_{\rm cx}$ assuming  eqs.(14) and (20). 
 In the gas, we use 
$\omega'= (\mu'-{\bar \mu})n'
 + (\mu_{\rm I}'-{\bar \mu}_{\rm I})n_{\rm I}'+ {\bar p}- p' $, 
where   the first  term  in the right hand side is negligible 
from  eq.(18).  Further we set 
$p'= p_{\rm cx}+ k_BT n_{\rm I}'$ from eq.(20)and use eq.(3)  to find   
\be 
 \omega'= k_B T n_{\rm I}' [\ln({n_{\rm I}'/{\bar n}_{\rm I}})   
   - 1-{\Delta {\nu}_s}] + {\bar p}-p_{\rm cx}.
\en
In the liquid,  ${\hat n}_{\rm I}$ is very small and  we obtain   
\be 
\hat{\omega}= k_B  T [{\hat n}_{\rm I} \ln
({\hat n}_{\rm I}/{{\bar n}_{\rm I}}) 
+\phi n_{\rm I}']  ,
\en 
where  ${\hat n}_{\rm I}= {{\bar n}_{\rm I}}
- \phi  {n}_{\rm I}'$ from eq.(13). 
Thus, if  $\phi \ll {\bar n}_{\rm I}/ n_{\rm I}'\ll 1$, 
the logarithm $\ln({\hat n}_{\rm I}/{{\bar n}_{\rm I}})$ 
can be  expanded   with respect to $\phi$, leading to $\hat{\omega}
\propto \phi^2$.  However,  we are also interested in  the case 
 $\phi \sim {\bar n}_{\rm I}/ n_{\rm I}'$.

In  equilibrium, $\Delta G$ in eq.(34)  is minimized with respect 
to $n_{\rm I}'$ and $\phi$, where $R$ and $S$ are functions of $\phi$ at fixed 
$\theta$ and $n_{\rm b}$. Thus, let us change $n_{\rm I}'$ and $\phi$ 
infinitesimally by $\delta n_{\rm I}'$ and $\delta \phi$, respectively. 
From eqs.(34)-(37), we calculate  
the incremental change of  $\Delta G$  as 
\bea 
\delta(\Delta G)&=& Vk_BT[ \ln(n_{\rm I}'/\hat{n}_{\rm I})-\Delta\nu_{\rm s}] 
(n_{\rm I}'\delta\phi+ \phi\delta n_{\rm I}') \nonumber\\
&&+ V [{\bar p}- p_{\rm cx}  +2\sigma/R-k_B T n'_{\rm I}]\delta\phi. 
\ena  
Therefore,  the equilibrium conditions (14) and (21) follow from 
$\p (\Delta G)/\p n_{\rm I}'= \p (\Delta G)/\p \phi=0$.

\begin{figure}
\includegraphics[width=0.91\linewidth]{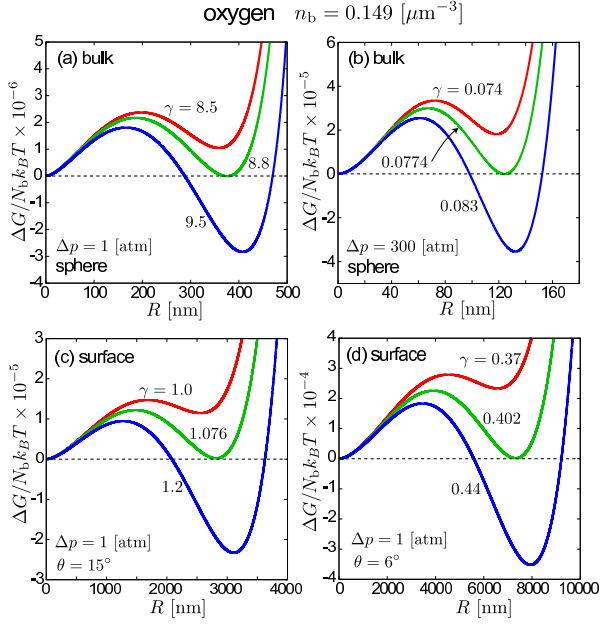}
\caption{(Color online)
Normalized  bubble free energy 
$\Delta G/N_{\rm b} k_BT$ vs $R$ for  O$_2$ in liquid  water 
 in the isobaric condition 
for  bulk bubbles in  (a) and (b) 
and for surface bubbles in (c) and (d), where  
 $T=300$ K  and $n_{\rm b}=N_{\rm b}/V=  0.149/$$\mu$m$^{3}$.
The pressure difference  $\Delta p= 
{\bar p}-p_{\rm cx}$ is 1 atm 
in (a), (c), and (d) and is 300 atm in (b). 
The contact angle $\theta$ is  $15^\circ$ 
in (c) and is  $6^\circ$ in (d). 
Then, ($A, \gamma_{\rm m})=$
$(3.86,7.74)$ in (a), $(0.0129,0.0680)$ in (b), 
$(0.367,0.945)$ in (c), 
and $(0.109,0.353)$ in (d), 
while    $\gamma_{\rm tr}\cong  1.14\gamma_{\rm m}$ for all 
the cases.   In  the equilibrium state at $R=R_2$  with 
$\Delta G_{\rm min}<0$ on lowest  curve in each panel, 
$(\gamma, \phi,R_2)=$   
 $(9.5, 4.25\times 10^{-2},0.41)$ in (a), 
 $(0.083, 1.45\times 10^{-3},0.13)$ in (b), 
 $(1.2, 1.62\times 10^{-2},3.1)$ in (c), and 
 $(0.44, 7.00\times 10^{-3},7.9)$ in (d) 
with $R_2$ in $\mu$m. 
}
\end{figure}

\begin{figure}
\includegraphics[width=1\linewidth]{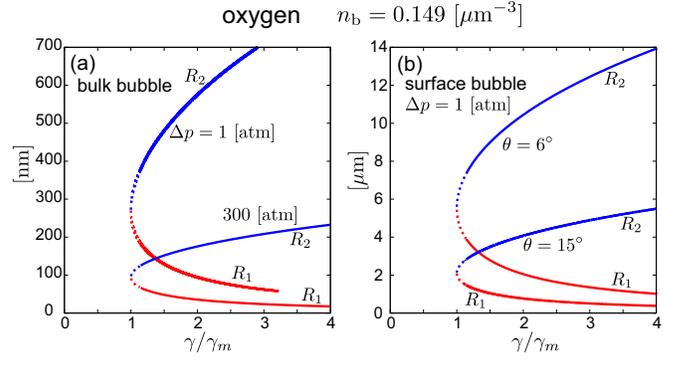}
\caption{(Color online) 
Two radii $R_1$ and $R_2$ vs $\gamma/\gamma_{\rm m}$  
giving the local maximum and minimum of $\Delta G$ 
for   O$_2$ in water at $T=300$ K in the isobaric condition. 
They are written in dotted lines in the 
 region $\gamma_{\rm m}<\gamma<\gamma_{\rm tr}$. 
As in Fig.2, displayed curves 
are for bulk bubbles with  $\Delta p= 
{\bar p}-p_{\rm cx}$ being  1 or  300 atm (left) 
or for surface bubbles with $\theta= 15^\circ$ or $6^\circ$ (right). }
\end{figure} 

\begin{figure}
\includegraphics[width=1\linewidth]{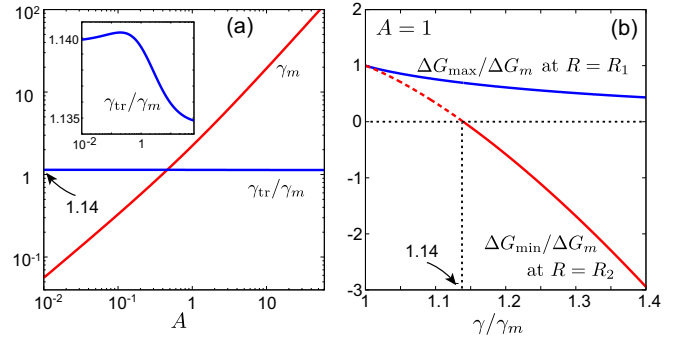}
\caption{(Color online) 
(a)  $\gamma_{\rm m}(A)$ and ratio 
$\gamma_{\rm tr}(A)/\gamma_{\rm m}(A)$ vs $A$ in the isobaric condition, 
where    $\Delta G(R,\gamma,A)$ as a function of $R$ 
exhibits local extrema 
for $\gamma>\gamma_{\rm m}$ 
and its local  minimum becomes negative 
for $\gamma> \gamma_{\rm tr}$. 
 Here, $\gamma_{\rm tr}/\gamma_{\rm m}\cong 1.14$ 
for any $A$. 
(b)  Local maximum $\Delta G_{\rm max}(\gamma,A)$  at $R=R_1$ 
and local minimum $\Delta G_{\rm min}(\gamma,A)$ at $R=R_2$ 
divided by  $\Delta G_{\rm m}(\gamma_{\rm m},A)$  
as functions of $\gamma/\gamma_{\rm m}$ for $A=1$. 
Here, $\Delta G_{\rm min}<0$ for $\gamma>\gamma_{\rm tr}$, where 
$R_2$ is the equilibrium bubble radius. 
}
\end{figure}

Furthermore, if we assume   the pressure balance (21) 
(without assuming eq.(14)), 
 $\Delta G$ becomes  a function of 
$R$ only under eqs.(28) and (35). 
Its derivative with respect to $R$  is calculated as 
\be 
\frac{d  (\Delta G)}{dR}= S \bigg[ \ln\bigg(\frac{n_{\rm I}'}{\hat{n}_{\rm I}}
\bigg)
-\Delta\nu_{\rm s}\bigg]\bigg 
[{\bar p}- p_{\rm cx}  +\frac{4\sigma}{3R}\bigg]
\en 
The extremum condition ${d  (\Delta G)}/{dR}=0$  gives 
$n_{\rm I}' ={\hat n}_{\rm I} e^{\Delta\nu_{\rm s}}$, leading to  
eqs.(14) and (15).

\subsection{Local maximum and minimum  of  bubble free energy 
at fixed pressure}

In Fig.2, we plot $\Delta G/N_{\rm b}k_BT$ vs $R$ for 
O$_2$ for bulk and surface bubbles 
in water  under eq.(21), where  $T=300$ K and ${\bar p}-p_{\rm cx}
=1$ or 300 atm.  When we use O$_2$ (in Figs.2, 3, 6, and 7), 
we fix the bubble density 
 at  $n_{\rm b}=N_{\rm b}/V=  0.149/$$\mu$m$^{3}$.
We recognize that  $\Delta G$  assumes a local maximum at $R=R_1$ 
and a negative 
minimum at $R=R_2$ for sufficiently large ${\bar n}_{\rm I}$. Therefore,  
bubbles can appear in equilibrium at $R=R_2$  
with increasing ${\bar n}_{\rm I}$.  
For the case ${\bar p}-p_{\rm cx}=$ 300 atm, 
the  pressure in the bubble interior 
 is also nearly equal to 300 atm for $R\gg 5$ nm  
and   the interior  oxygen density 
is $n_{\rm I}'= 7.2$ $/$nm$^3$ from 
the ideal-gas formula ($p=nk_BT$). 
Instead, if we use  the van der Waals 
equation of state    
($p =nk_BT/(1-n/3n_c) - (9k_BT_c/8n_c)  n^2$)  
at $p=300$ atm and $T=300$ K, 
 the density becomes 7.9 $/$nm$^3$, where 
$T_c= 154.6$ K 
and $n_c= 8.0$ nm$^{-3}$ for O$_2$. 
Thus, the van der Waals  interaction 
among O$_2$ molecules is smaller than 
$k_BT $ (per molecule) even  at 
${\bar p}-p_{\rm cx}=300$ atm.

To explain Fig.2, we treat $\Delta G$ as a function of $R$  
by  increasing   ${\bar n}_{\rm I}$ or $\gamma$ 
with the other parameters fixed. 
As will be shown  in Appendix B, 
 $\Delta G$ monotonically increases 
for $\gamma<\gamma_{\rm m}$  and   
 exhibits  a local maximum $\Delta G_{\rm max}$ 
at $R= {R}_{1}$ and a local minimum $\Delta G_{\rm min}$ at  $R=R_2$, where 
  $\Delta G_{\rm min}>0$ for $\gamma_{\rm m}<\gamma<\gamma_{\rm tr}$  
and  $\Delta G_{\rm min}<0$  for $\gamma>\gamma_{\rm tr}$.  Here, $R_2/R_1$ 
increases from 1 with increasing $\gamma$ above $\gamma_{\rm m}$. 
In each panel in  Fig.2, we set $\gamma_{\rm m}<\gamma<\gamma_{\rm tr}$ for 
the upper curve, $\gamma=\gamma_{\rm tr}$ for the middle curve, and 
$\gamma>\gamma_{\rm tr}$ for the lower curve.  
Therefore, the two-phase states at $R=R_2$ 
are metastable for $\gamma_{\rm m}<\gamma<\gamma_{\rm tr}$ 
and stable for  $\gamma\ge \gamma_{\rm tr}$.
In Fig.3, we plot $R_1$ and $R_2$ 
for  O$_2$ in water for bulk and  surface bubbles.

In Appendix B,  we shall see that  $\gamma_{\rm m}$ and 
$\gamma_{\rm tr}$  depend only 
on the following dimensionless parameter, 
\be 
A= [2\sigma/({\bar p}-p_{\rm cx})][4\pi G(\theta)n_{\rm b} e^{\Delta\nu_{\rm s}}/3]^{1/3} ,
\en 
which  diverges  as ${\bar p}\to p_{\rm cx}$ 
and becomes small  with increasing 
 ${\bar p}-p_{\rm cx}$ and/or  decreasing $G(\theta)n_{\rm b}$. 
Using  $A$, we may rewrite eq.(26)  in terms of $u=R/R_c$ as 
\be 
A^3u^3=\gamma^4 (u-1)/(u+\gamma),
\en 
which  holds  for $u=R_1/R_c$ and $R_2/R_c$. 
In  Fig.4, we plot  $\gamma_{\rm m}$ and $\gamma_{\rm tr}$ vs $A$ in (a) 
 and display  $\Delta G_{\rm max}$ 
and $\Delta G_{\rm min}$ as functions of 
$\gamma/\gamma_{\rm m}$ 
 at $A=1$ in (b).
Here,  $\gamma_{\rm m} \sim A^{3/4}$ for $A\ll 1$  and  
  $\gamma_{\rm m} \sim A$ for $A\gg 1$. 
For any $A$, we find  
\be 
\gamma_{\rm tr}/\gamma_{\rm m}\cong 1.14.
\en 
For example,  $\gamma_{\rm m}=0.330, 2.23,$ and 19.4 
for $A=0.1$, 1, and 10, respectively. 
The threshold of bubble formation $(\gamma>\gamma_{\rm tr})$ 
is thus approximately given by 
${\bar n}_{\rm I} \gs {n}_{\rm I}^c$  for $A\ls 1$ 
and by ${\bar n}_{\rm I} \gs A{n}_{\rm I}^c$  for $A\gs 1$.


In particular, with increasing the solute density, 
we examine  the case 
 $ \gamma\gg \gamma_{\rm m}\sim \gamma_{\rm tr}$ using eq.(41), 
where   $R_2$ is the equilibrium radius. 
Then, for any $A$, we find  
\be 
R_1 / R_c \cong 1+ A^3(1+\gamma)/\gamma^4+\cdots \cong 1 . 
\en 
The ratio  $R_1/R_c $ rapidly approaches $1$ with increasing $\gamma$. 
  In fact,  even  at $\gamma=\gamma_{\rm tr}$, 
we have  $R_1/R_c= 1.091$, 1.122, and 1.150  
for  $A=0.1$, 1, and 10, respectively. 
On the other hand, supposing $u=R_2/R_c \gg 1$, 
 we obtain  $u^2(u+\gamma) \cong \gamma^4 /A^3$ from eq.(41). 
 For $A\ll 1$, we have 
\be 
R_2/R_c \cong  \gamma^{4/3}/A \gg 1.
\en 
For $A \gg 1$, there are two limiting cases:
\bea 
R_2/R_c  &\cong&  (\gamma/A)^{3/2}\gg 1 \quad (A\ll \gamma \ll A^3)\\
&\cong&   \gamma^{4/3}/A  \gg \gamma \quad \quad (\gamma \gg A^3).
\ena 
In these limiting cases, we surely obtain $R_2\gg R_c$. 
See Fig.10  in Appendix B 
for  the behaviors of  $R_1$ and $R_2$ vs $\gamma$. 
On the other hand, for $ \gamma\gg \gamma_{\rm m}$, 
 the solute fraction $\alpha$ in bubbles 
in eq.(16) is much smaller than 1  at  $R=R_1$ and 
approaches 1 at  $R=R_2$ (see Fig.10(b)).

\subsection{Bubble free energy at fixed chemical potentials}

So far we have fixed the total particle numbers 
$N= V( {\hat n}+\phi n')$ and 
$N_{\rm I}= V( {\hat n}_{\rm I}+\phi n_{\rm I}')$ as well as the 
liquid pressure. In this case,  
 the water chemical potential $\hat\mu$ 
is nearly fixed at the initial value ${\bar\mu}$  
from eq.(18). As another boundary condition, 
we may  attach a solute reservoir 
to the cell to fix the solute 
chemical potential at the initial value 
${\bar\mu}_{\rm I}$, where we still  attach  a pressure valve. 
In this grand canonical case, 
we have  ${\hat n}_{\rm I}={\bar n}_{\rm I}$ so that 
${\hat \omega}=0$ from eq.(32).   We should 
 minimize the grand potential, 
\bea 
\Omega &=& G- {\bar\mu}N-{\bar\mu}_{\rm I}N_{\rm I}\nonumber\\ 
&=&  V \phi\omega' +\sigma S,
\ena 
where $G$ is defined in eq.(31),  
${\bar\mu}$ and ${\bar\mu}_{\rm I}$ are the initial chemical 
potentials,  and $\omega'$ is given by eq.(36). 
To derive  the second line of eq.(47), we have used 
the relation  ${F}/{V} = 
\phi\omega'- (1+\phi){\bar p}
+ {\bar \mu}(\hat{n}+ \phi n')
+{\bar \mu}_{\rm I}({\hat{n}}_{\rm I}+ \phi n_{\rm I}')$,  
where  $N$ and $N_{\rm I}$ are not fixed.

With respect to small changes $n_{\rm I}'
\to  n_{\rm I}'+\delta n_{\rm I}'$ and 
$\phi\to \phi+\delta\phi$, 
the incremental change of  $\Omega $  is 
given by the right hand side of eq.(38)  if ${\hat n}_{\rm I}$ 
 is replaced by ${\bar n}_{\rm I}$. 
Therefore,  the extremum conditions  
$\p \Omega/\p n_{\rm I}'= \p \Omega /\p \phi=0$ yield 
the pressure balance (21) and the chemical equilibrium condition 
${n}'_{\rm I}={\bar n}_{\rm I}e^{\Delta\nu_{\rm s}}$.
If these extremum conditions are 
assumed, we obtain 
\be 
\omega'= -   ({\bar p}-p_{\rm cx})\gamma ,
\en    
which is negative for $\gamma>0$ outside CX. 
Here,  in the second line of eq.(47), the first term  
 is proportional to $ R^3$ and  the second term to $ R^2$, 
so the  minimum of $\Omega$ 
 decreases monotonically  with increasing $R$  (for $R>R_c$), 
 indicating  appearance of  macroscopic bubbles.

\section{ Solute-induced  nucleation}
\subsection{Experimental situations} 

We have shown that the bubble free energy 
$\Delta G(R)$ has a local maximum at $R=R_1$ 
and a minimum at $R=R_2$ for  $\gamma>\gamma_{\rm tr}$ 
(except for gas films). In such situations, 
the initial homogeneous  state is metastable and  
there can  be   solute-induced bubble  nucleation 
outside CX   
\cite{Onukibook}. In contrast,  in one-component fluids, bubble nucleation  
 occurs only inside CX (${\bar p}<p_{\rm cx}$)
\cite{Katz,Caupin,Binder,hetero,Onukibook}. 
In nucleation, crucial is the   free energy $F_c$ 
needed to create a critical bubble 
with    $R=R_c$.  We  call it the 
 nucleation barrier, since the nucleation rate $I$ of bubble formation 
is proportional to the Boltzmann factor $\exp(-F_c/k_BT)$. 
Therefore,  if $F_c/k_BT$ 
is too large (say, 80), $I$  becomes too small for experiments 
on realistic  timescales.   
In our case, $F_c$ is reduced with 
increasing ${\bar p}-p_{\rm cx}$ 
and/or  $ \gamma ={\bar n}_{\rm I}/{ n}_{\rm I}^c-1$. 
For surface bubbles, 
it is  also  reduced  with decreasing the contact angle $\theta$.

We make some comments on experimental situations.  
First, in the previous observations 
 \cite{Bunkin,ex2,ex3,ex4},    bulk nanobubbles 
have been produced by breakup of 
large bubbles composed of a 
 gas such as O$_2$, CH$_4$, or Ar, 
where the typical flow-induced 
bubble size  is of great interest 
\cite{Onukishear}. 
Second,  a small amount of surfactants and/or  ions 
  are usually  present in water, which 
 increase  the bubble stability  \cite{review2}. 
Indeed, surfactant molecules at the gas-liquid interface reduce 
the surface  tension, while  electric charges or 
 electric double layers  
at the interface   prevent  bubble coalescence
\cite{salt,salt1,Grac,Takahashi}. 
For example, with addition of   O$_2$ and a salt  in water\cite{ex3}, 
the bubble-size distribution on long timescales 
was found to have a  peak at $R\sim 100$ nm. 
Third,  on a non-smooth hydrophobic wall, 
there can be preexisting trapped bubbles  or 
 strongly  hydrophobic spots. 
In such  cases, 
there should be no significant nucleation barrier for 
the formation of surface bubbles 
 with  small contact angles $\theta$.

\subsection{Critical radius and nucleation barrier }

\begin{figure}
\includegraphics[width=1\linewidth]{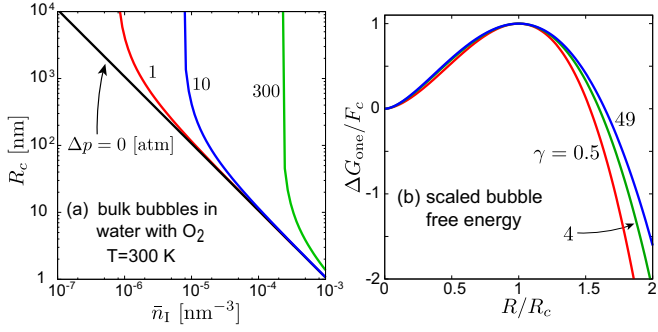}
\caption{(Color online) (a)  Critical  radius $R_c$ 
in eq.(27)  vs oxygen density $ {\bar n}_{\rm I}$ 
for bulk bubbles  in water, 
where $\Delta p= 
{\bar p}-p_{\rm cx}= 0, 10$, and 300 atom and  $T=300$ K. 
  (b)  $\Delta G_{\rm one} (R)/F_c$ in eq.(51) vs $R/R_c$ 
for $\gamma =0.5,4,$ and 49.     }
\end{figure}
\begin{figure}
\includegraphics[width=1\linewidth]{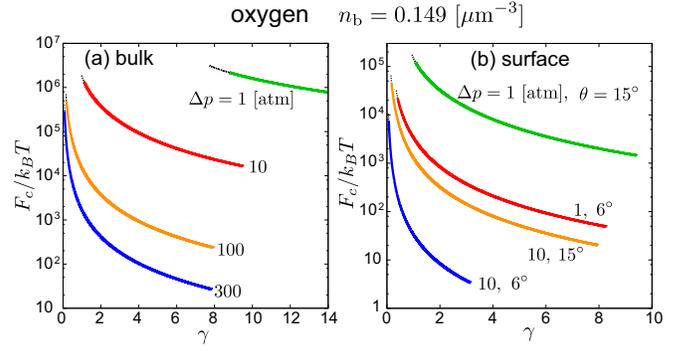}
\caption{(Color online) 
Normalized nucleation barrier $F_c/k_BT$ 
vs $\gamma$ for O$_2$ in water.
for (a) bulk and (b) surface bubbles.  
}
\end{figure}

We consider a single  bubble with curvature  $R^{-1}$ 
in bulk or on a hydrophobic wall.  
In the early stage with small  $\phi$,  
we may neglect $\hat{\omega}\propto \phi^2$ in eq.(34)  
  to obtain  the  single-bubble free energy in the standard form  
\cite{Caupin,Binder,hetero,Onukibook,Katz}, 
\be 
 \Delta G_{\rm one}(R) 
 = G(\theta)\bigg [  \frac{4}{3}\pi R^3\omega' +4\pi \sigma R^2 
\bigg ] , 
\en  
where $\omega'$ is given by eq.(36)  and 
   $n_{\rm I}'$ is related to $R$ by  
the pressure balance (21). 
 Note that $\omega'$ 
 is usually a negative constant in nucleation in metastable systems.  
Here,   $d(\Delta G_{\rm one})/dR\propto \ln(n_{\rm I}'/{\bar n}_{\rm I})
-\Delta\nu_{\rm s}$, which follows from  
eq.(39) if ${\hat n}_{\rm I}$ is replaced by 
${\bar n}_{\rm I}$.  Then,  $  \Delta G_{\rm one}(R) 
$ is maximized at  
 the critical radius $R_c$ in eq.(27). 
See  Fig.5(a) for  $R_c$ vs ${\bar n}_{\rm I}$ 
for O$_2$.  Since $\omega'= -2\sigma/R_c$ at $R=R_c$ from eq.(36), 
the nucleation barrier ($=$the maximum of $\Delta G_{\rm one}$ at $R=R_c$) 
 is written as    
\be 
F_c= 4\pi G(\theta) \sigma R_c^2/3=16\pi G(\theta) 
\sigma^3/[( {\bar p}-p_{\rm cx})\gamma]^2.
\en 
For surface bubbles 
with small $\theta$, we have   $G(\theta)\cong 3\theta^4/16$ 
and $G_c\propto \theta^4$. 
For bubble nucleation in one-component fluids, 
 $F_c$  is given by the above form 
with  $\gamma=1$ and $ {\bar p}<p_{\rm cx}$. 
In terms of   $u=R/R_c$, we may also 
express $\Delta  F_{\rm one}(R)$ simply as  
\be 
 \Delta G_{\rm one}/F_c= 
2u^2 (u/\gamma+1)\ln \bigg(\frac{1+\gamma/u}{1+\gamma}\bigg)+ u^2.
\en 
The right hand side may be approximated by 
  $ -2u^3+ 3u^2$ for $\gamma\ll 1$ 
and  by $  -2u^2 \ln u+ u^2$ for $\gamma\gg 1$.
In Fig.5(b), we plot the above scaling function.  
In addition,  the nucleation  rate $I$ is of the form \cite{Onukibook}
\be 
I= \Gamma_c n_{\rm cx}^{\ell}  \exp( -F_c/k_BT), 
\en 
where  $\Gamma_c$ 
is  the growth rate  of a critical bubble (see eq.(60)  
in the next subsection).

For  water at $T=300$ K,  we have 
\be
F_c/k_B T \cong 73 G(\theta) R_c^2,
\en 
with $R$ in nm.  In   Fig.6,  we plot $F_c/k_BT$ vs ${\gamma}$ 
for O$_2$ in water for bulk and surface bubbles. .
In homogeneous bubble  nucleation of pure water at $T\sim 300$ K 
\cite{Katz,Caupin}, bubbles with $R>R_c$ are 
detectable for $F_c/k_BT \ls 70$ or  
for $R_c \ls 1$ nm in experimental times and 
$R_c $ can be of order   1 nm  only for 
negative ${\bar p}$ of order $ -1000$ atom. 
For   O$_2$ in our case,  $R_c$ is  decreased  
down to  1 nm,  depending on 
 ${\bar n}_{\rm I}$, ${\bar p}-p_{\rm cx}$, 
and $\theta$ in Fig.6.

\subsection{Dynamics of bulk nucleation}
\begin{figure}
\includegraphics[width=1\linewidth]{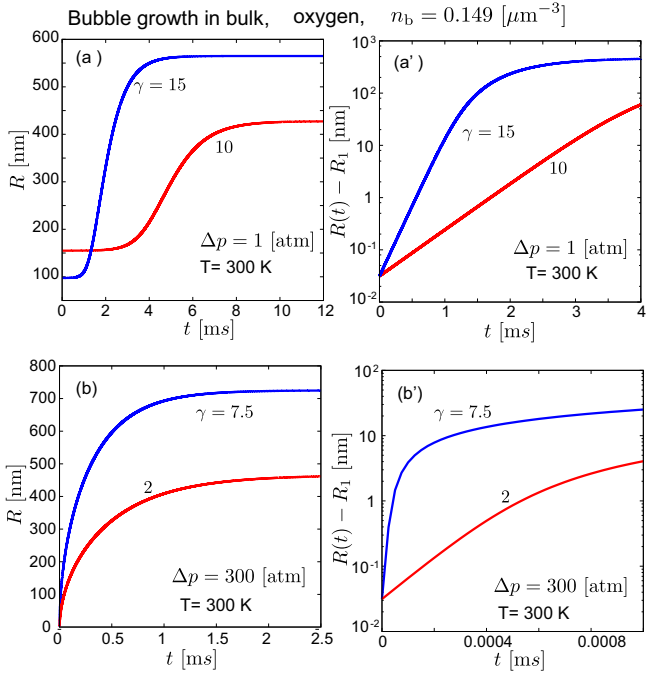}
\caption{(Color online) 
Growth of  radius $R(t)$ 
for bulk bubbles in the isobaric condition 
 for O$_2$ in water at $T=300$ K, 
 which are obtained from  eq.(59) 
for  $R(0) = R_1+ 0.3$ ${\rm \AA}$   
with $D=2\times 10^{-9}$ cm$^2/$s. 
In (a) and (a'), $\gamma=10$ and 15  
with  $\Delta p={\bar p}-p_{\rm cx}=$1 atom. 
In (b) and (b'), $\gamma=2$ and 7.5  
with  $\Delta p=$300 atm. 
Curves  are written on linear scales 
in (a) and (b)  and on semi-logarithmic scales in (a') and (b').  
Here, $\gamma_{\rm tr}=8.80$ and $0.0774$ for 
 $\Delta p=1$ and  $300$ atm, respectively.
}
\end{figure}

We next examine nucleation dynamics 
of bulk bubbles for $\gamma>\gamma_{\rm tr}$. To describe  
attainment of the equilibrium radius $R_2$ in the simplest manner, 
 we assume a common radius $R(t)$ for all the bubbles 
with a constant  bubble density $n_{\rm b}$. 
We also assume a time-dependent 
 background solute density in the liquid defined by  
\be 
\hat{n}_{\rm I}(t)= {\bar n}_{\rm I}- \phi(t)n_{\rm I}'(t),
\en  
where $\phi(t)$ is determined by eq.(28) 
 with $G(\theta)=1$. Expressing  
$\phi(t)$ and $n_{\rm I}'(t)$ in terms 
of  $R(t)$, we may describe saturation of $\phi(t)$ up to the 
 equilibrium volume fraction. 
After this stage, however,  the bubble number 
  decreases in time  in the presence of  bubble coalescence 
(which can be suppressed with addition of  salt\cite{salt,salt1}).

For simplicity, we further  assume that 
the solute diffusion constant $D$ 
is much  smaller   than  the thermal diffusion constant $D_T$ in 
the liquid. Then,    we can neglect temperature inhomogeneity 
 around   bubbles, which  much simplifies 
the calculation. 
In fact, for liquid water at $300$ K and 1 atom, 
we have $D_T \sim 1.4\times 10^{-3}$ cm$^2/$s 
and $D\sim 2.0 \times 10^{-5}$ cm$^2/$s ($\ll D_T$)  for O$_2$.

We focus our attention 
to  a single bubble neglecting its Brownian motion, where  
$n_{\rm I}(r,t) $ slowly changes in time $t$ 
tending to ${\hat n}_{\rm I}(t) $ in eq.(54) far 
from it. We write the distance from the droplet center as $r$.
In  the bubble exterior $r>R$, the solute obeys the diffusion equation,   
\be 
\frac{\p n_{\rm I}}{\p t} = D \nabla^2 n_{\rm I}. 
\en   
 We assume the continuity of the solute chemical potential $\mu_{\rm I}$ 
at $r=R+0$ and  $r=R-0$ across the interface.
From eq.(7), the solute density $n_{\rm I}^R= n_{\rm I}(R+0,t)$ 
immediately outside the bubble 
 is related to the interior density  $n_{\rm I}'$ by  
\be 
n_{\rm I}^R= n_{\rm I}' e^{-{\Delta  \nu}_s}.
\en 
Therefore, in the quasi-static approximation\cite{Onukibook},  
$n_{\rm I}(r,t)$ slightly outside 
 the interface is written as 
\be 
n_{\rm I}(r,t)=  {\hat  n}_{\rm I}+
({ n}_{\rm I}^R  - {\hat n}_{\rm I}) R/r .
 \en 
The flux to the bubble is given by $D ( {\hat n}_{\rm I}-{ n}_{\rm I}^R)/R$,  
so  the conservation of the solute  yields 
\be 
({ n}_{\rm I}'  - { n}_{\rm I}^R) \frac{d R}{d t}
= \frac{D}{R} ( {\hat n}_{\rm I}- n_{\rm I}^R  )
\en 
Here, ${\hat n}_{\rm I}- n_{\rm I}^R= 
{\bar n}_{\rm I}- n_{\rm I}'(\phi+ e^{-\Delta \nu_{\rm s}})$ from eqs.(13) 
and (55, so the right hand side of eq.(58) vanishes 
at  $R=R_2$ from eq.(13). In accord with 
the equilibrium relation (27), 
division of eq.(58) by $n_{\rm I}' D/R$  gives    the desired equation, 
\be 
(1-e^{-{\Delta  \nu}_s})\frac{ R}{D}  \frac{d R}{d t}
= \frac{ ({\bar n}_{\rm I}-n_{\rm I}^c )(1-R_c/R)}{(
{\bar p}- p_{\rm cx}+2\sigma/R)/k_BT} 
 -\phi.
\en

For $\gamma>\gamma_{\rm tr}$, 
the right hand side of eq.(59) vanishes 
for $R=R_1$ and $R_2$, where $R_1\cong R_c$. 
Here, bubbles with $R>R_1$ grow up to $R_2$, 
while those with $R<R_1$ shrink.  
If the deviation  $\delta R=R-R_1$ is small, it obeys  
the linear  equation 
 $d(\delta R)/dt= \Gamma_c\delta R$, where $\Gamma_c$ 
is  the growth rate of a critical bubble of the form,  
\be 
\Gamma_c= 
D R_c^{-2}\gamma/[(1+\gamma)( e^{\Delta \nu_{\rm s}}-1)]. 
\en 
In terms of $\Gamma_c$ and $u=R/R_c$, we may rewrite eq.(59) as  
\be 
 \frac{du}{dt}={\Gamma_c} \frac{1+\gamma}{u}\bigg[ 
\frac{u-1}{u+\gamma}
- \frac{A^3}{\gamma^4}  u^3 \bigg], 
\en 
which is consistent with eq.(41).

In Fig.7, we display the growth of $R(t)$ by 
setting $D=2\times 10^{-5}$ cm$^2/$s for O$_2$ in water at $T=300$ K, 
 where ${\bar p}-p_{\rm cx}$ is 1 atm in (a) and (a') 
and 300 atm in (b) and (b'). 
As the initial radius, we set $R(0)= R_1+ 0.3$ ${\rm \AA}$, 
which yields   $u(0)-R_1/R_c\sim 10^{-2}$ in eq.(61). 
The right panels indicate   the exponential growth, 
\be 
R(t)= R(0)+ (R(0)-R_1)e^{\Gamma_ct}
\en 
in the 
early  stage. Numerically,  $\Gamma_c$ is $ 2.89$ and $6.70$ 
for $\gamma=10$ and 15, respectively, 
in (a) and (a'), while it is $ 7.63$ and $142$ 
for $\gamma=2$ and 7.5, respectively, 
in (b) and (b'). These  values  agree with eq.(59). 
In this calculation, we assume the pre-existence 
of bubbles with radii slightly 
exceeding $R_1$. However, if we start with the 
homogeneous initial state, the  birth of such large bubbles in the cell 
occurs as rare thermal activations 
on a  timescale  
of order,
\be 
1/VI 
\sim  \exp({F_c/k_BT})/Vn_{\rm cx}^\ell \Gamma_c.
\en

\section{Summary} 

We have investigated bubble formation 
in bulk and on hydrophobic walls 
  induced by  accumulation of  a small amount of 
a neutral  solute in liquid water 
 outside the solvent  CX.  We have used  the fact that a gas such as 
O$_2$ or N$_2$  remains in gaseous states 
within  phase-separated domains in ambient 
liquid water, because it is mildly  hydrophobic 
    with a critical temperature much 
below 300 K. 
With this input, we  have 
constructed a simple thermodynamic theory 
for dilute binary mixtures including   a 
considerably large solvation chemical 
potential difference $\Delta \mu_{\rm s}=k_BT\Delta\nu_{\rm s} $. 
We have assumed  fixed particle numbers  
and a fixed liquid pressure 
 $(N$-$N_{\rm I}$-$p)$  in the text and in Appendix B, but  we 
have also treated bubble formation in the  
$\mu$-$\mu_{\rm I}$-$p$   ensemble in Sec.IIIC and 
in the $N$-$N_{\rm I}$-$V$ ensemble 
in Appendices  A and B,

In particular, in Sec.II, 
we have found a threshold solute density $n_{\rm I}^c$ 
in eq.(23) for film formation on a completely dried wall 
at fixed pressure, 
which is very  small for large $\Delta\nu_{\rm s}$. 
The threshold density is increased to 
$(\gamma_{\rm m}+1) n_{\rm I}^c$  for metastable bubbles 
and $(\gamma_{\rm tr}+1) n_{\rm I}^c$  for stable bubbles due to 
the surface tension, where $\gamma_{\rm tr}\cong 1.14 \gamma_{\rm m}$. 
Here,  $\gamma_{\rm m}$ and  $\gamma_{\rm tr}$ 
are displayed in Fig.4(a) as functions of 
a parameter $A$ in eq.(40). In Sec.III, 
we have also presented a bubble  free energy $\Delta G$ 
for a small gas fraction $\phi$  in eqs.(34)-(37) for 
the isobaric case, whose minimization yields 
the equilibrium conditions (14) and (21).
In Sec,IV, we have calculated the critical radius $R_c$ and 
the   barrier free energy $F_c$ for solute-induced 
nucleation.  The $F_c/k_BT$  is   
very high for homogeneous  nucleation  
except for  high liquid pressures,  
but it can be decreased for heterogeneous nucleation 
with a small contact angle $\theta$.

We make some critical remarks. (i) First, we have assumed 
gaseous domains. 
However, with increasing ${\bar p}-p_{\rm cx}$ and/or  $\Delta\nu_{\rm s}$, 
  liquid or solid precipitates should be formed  
in bulk and on walls, sensitively depending on their 
mutual attractive interaction.  Note that a large attractive 
interaction arises even among large hard-sphere particles  
in water due to deformations of the hydrogen bonding 
 \cite{Paul,Garde,Chandler,Pratt}.   (ii) Second, we should include  the 
effects of surfactants and ions in 
the discussion of the bubble size distribution 
 \cite{review2,salt,salt1,Grac,Takahashi}.  
 In this paper, we have assumed a constant 
 bubble density $n_{\rm b}$ in eqs.(28), (35), and (40). 
This assumption  can be justified only when  bubble coalescence 
is suppressed by the electrostatic interaction 
 near the gas-liquid interfaces. 
(iii) Third,  dynamics of 
bubble formation and dissolution\cite{Teshi}  
should be studied in future,   which can be induced by a 
change in pressure,  temperature, or solute density.

\acknowledgments
This work was supported by KAKENHI No.25610122. 
R.O. acknowledges support from the Grant-in-Aid for Scientific Research on
Innovative Areas ``Fluctuation and Structure" from the Ministry of Education,
Culture, Sports, Science, and Technology of Japan.

\vspace{2mm}
\noindent{\bf Appendix A: 
Bubbles  at fixed cell volume}\\
\setcounter{equation}{0}
\renewcommand{\theequation}{A\arabic{equation}}

Here, we consider  two-phase coexistence 
outside CX, fixing the particle numbers and 
the cell volume $V$ without  a pressure valve. 
Some discussions were already made 
on attainment of two-phase equilibrium 
in finite systems inside CX (without 
impurities) \cite{Onukibook,Binder1}.

In this case, while eqs.(14) and (15) are unchanged, 
  the liquid volume is decreased  by $\phi V$ 
 for  $n'\ll {\bar n}$ with appearance of a gas region. 
As a result,   the liquid  density 
$\hat n$ is increased   as  
\be 
{\hat n}= (1+\phi){\bar n}.
\en  
In terms of the isothermal compressibility $K_T$ 
of liquid water, the pressure increase 
is given by $\phi/K_T$, so the pressure balance relation 
(21) is changed as 
\be 
k_BT n_{\rm I}'= {\bar p}-p_{\rm cx} + \phi/K_T + 2\sigma/R.
\en 
Here,    $K_T =0.45 \times 10^{-3}/$MPa  in ambient water near CX, where 
even a very small $\phi$ gives rise to a large pressure.

The bubble free energy  $\Delta F$
 is defined as  
the increase in the Helmholtz free energy as 
$F={\bar F}+ \Delta F$ due to appearance of bubbles.  
Some calculations give 
\be
\Delta F =V[\phi\omega'+ (1-\phi) \hat{\omega}] +\sigma S .
\en 
Here, we may replace $(1-\phi) \hat{\omega}$ 
by $ \hat{\omega}$ for small $\phi$. 
Then $\Delta F$  assumes the form of eq.(34), 
but we need to change $\hat{\omega}$ in eq.(37) as     
\be 
\hat{\omega}=  T [{\hat n}_{\rm I} \ln
({\hat n}_{\rm I}/{{\bar n}_{\rm I}}) 
+\phi n_{\rm I}']  + {\phi^2}/{2K_T}, 
\en 
where  the last term is due to the compression in the liquid. 
From  eq.(1), 
it is equal to $f_{\rm w}''({\bar n}) (\hat{n}- {\bar n})^2/2$, 
where  $f_{\rm w}''=\p^2 f_{\rm w}/\p n^2= 1/n^2K_T$.  
The counterpart of eq.(38) for the increment $\delta (\Delta F)$ is obtained if ${\bar p}-p_{\rm cx}$ 
is replaced by  ${\bar p}-p_{\rm cx}+\phi/K_T$. 
Minimization of $\Delta F$ with respect to $\phi$ and $n_{\rm I}'$ 
thus  yields eqs.(14) and (A2). Furthermore, 
 the derivative 
 $ d  (\Delta F)/dR $ is obtained if 
 ${\bar p}-p_{\rm cx}$ is replaced by $ 
{\bar p}-p_{\rm cx}+2\phi/K_T$ in the right hand side of  
 eq.(39).

The  equilibrium equation for $R$ or $\phi$ 
is given by eq.(22) if ${\bar p}- p_{\rm cx} $ 
is replaced by ${\bar p}- p_{\rm cx} +\phi/K_T$. 
In particular, for a gas film $(R^{-1}=0$), $\phi$ is 
explicitly calculated as  
\be 
{\phi}= e^{-\Delta{\nu}_s} 
\bigg[\sqrt{(1+h)^2/4+ \gamma h}- \frac{1+h}{2}\bigg], 
\en 
where we define 
\be 
h =K_T ({\bar p}- p_{\rm cx}) e^{\Delta\nu_{\rm s}} .
\en
For O$_2$ in ambient water, 
we have  $h=1.4\times 10^{-3}({\bar p}- p_{\rm cx})$ 
with pressures in atm, so  $h\ll 1$ for ${\bar p}\ll 10^3$ atm. 

Here, we assume $1+h>0$. A film appears for $\gamma>0$ as in the 
isobaric case.  In particular, for
    $\gamma |h| \ll (1+h)^2/4$, 
we find the linear behavior $\propto \gamma$ as     
\bea 
\phi&\cong&  \gamma e^{-\Delta\nu_{\rm s}}{h}/({1+h}) \nonumber\\
&\cong&   \frac{K_T}{1+h} 
\bigg[k_BT {\bar n}_{\rm I}e^{\Delta\nu_{\rm s}} -{\bar p}+ p_{\rm cx}\bigg]. 
\ena 
From  the first line, this formula  tends to eq.(25) only for $h\gg 1$. 
The second line can be used  even for ${\bar p}\le  p_{\rm cx}$, where  
 $\phi$ increases  with increasing  $ p_{\rm cx}-{\bar p}$ 
and/or ${\bar n}_{\rm I}$. 
See Appendix B for more analysis 
for the case $R^{-1}>0$.

\vspace{2mm}
\noindent{\bf Appendix B: 
Scaling of bubble free energy}\\
\setcounter{equation}{0}
\renewcommand{\theequation}{B\arabic{equation}}
\vspace{1mm}

Here, we examine   the bubble free energy $\Delta F$  
in eq.(34) by scaling it in a  dimensionless form,
 assuming a common curvature $R^{-1}$ for all the bubbles.

\vspace{2mm}
{\it  1. Fixed pressure}\\
At fixed pressure in Fig.1, 
we assume  the pressure balance (21) 
and  introduce  scaling  variables $s$ and $v$ by  
\bea 
s &=& \phi e^{\Delta\nu_{\rm s}}= (4\pi  G(\theta)n_{\rm b}/3) e^{\Delta\nu_{\rm s}}R^3,\\
 v&=& [2\sigma /({\bar p}-p_{\rm cx})R]= A s^{-1/3}.
\ena
where  $A$ is the parameter  in eq.(40). 
From eq.(16) the solute fraction 
in bubbles is $\alpha=s/(1+s)$. 
 As a scaled bubble free energy,  we  define $\cal F$  as  
\be 
{\cal F}=  e^{\Delta\nu_{\rm s}}\Delta G /[V({\bar p}-p_{\rm cx})].
\en 
From eqs.(34)-(37), we express $\cal F$ in terms of $s$ and $v$ as  
\bea 
&&\hspace{-6mm}
{\cal F}
=  [\gamma+1-s(1+{v}) ] \ln\bigg[1-s\frac{1+{v}}{1+\gamma}\bigg ]
 \nonumber\\
&&\hspace{-2mm} +s(1+ {v}) \ln \bigg[\frac{1+{v}}{1+\gamma}\bigg]
+ + \frac{3}{2}sv +s ,
\ena  
where $\gamma$ is given by  eq.(24).  With fixed  $\gamma$ and $A$, $\cal F$ is a function of  $s$ only. 
From eq.(B2) its derivative with respect to $s$ is calculated as 
\be 
\frac{\p}{\p s}{\cal F}= 
\bigg(1+ \frac{2}{3}v\bigg)\ln \bigg[\frac{1+v}{\gamma+1-s(1+v)}\bigg ].
\en 
The extremum  condition ${\p}{\cal F}/\p s=0$ yields 
\be 
\gamma=s+ (1+s)v= s+ A (1+s)s^{-1/3} ,
\en 
which is equivalent to  eqs.(21), (26), and (41). 
If eq.(B6) is assumed, we have ${\cal F}= 
(3v/2+ 1)s - (1+\gamma)\ln(1+s)$ as extremum values depending only on $A$.
In Fig.8, we display ${\cal F}(\gamma, s, A)$ in the $\gamma$-$s$ plane 
at $A=1$.

\begin{figure}
\includegraphics[width=0.8\linewidth]{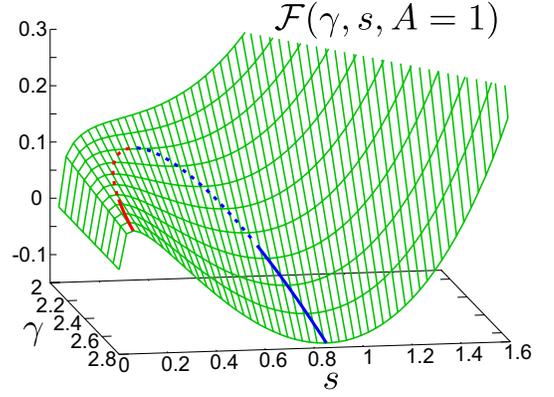}
\caption{(Color online) 
${\cal  F}$ in the isobaric condition 
in the  $\gamma$-$s$  plane 
at $A=1$ in the neighborhood of $\gamma=\gamma_{\rm m}$ and $s=s_{\rm m}$.   
Curve of $(\p {\cal F}/\p s)_{\gamma}=0$ or eq.(B6) 
is written on the surface, on which $s=s_1(\gamma,A)$ 
for the local maximum and $s=s_2(\gamma,A)$ 
for the local minimum for each $\gamma$. 
}
\end{figure} 

\begin{figure}
\includegraphics[width=1\linewidth]{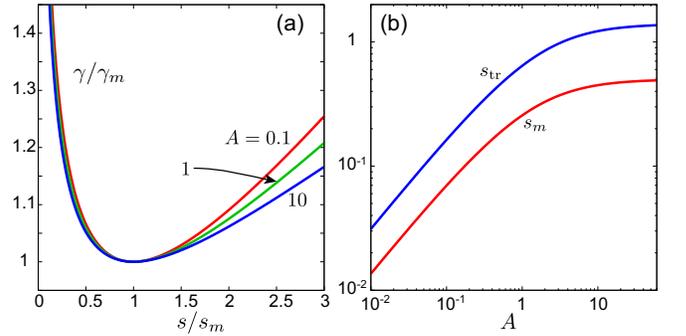}
\caption{(Color online)  Results in the isobaric condition.  
(a)  $\gamma/\gamma_{\rm m}$ vs $s/s_{\rm m}$ from 
the extremum condition (B6)    
for $A=0.1$, 1, and 10 and   (b)  $s_{\rm m}$ and $s_{\rm tr}$ vs $A$. 
Here,  $\gamma_{\rm m}$, $s_{\rm m}$, and $s_{\rm tr}$ are 
 determined by eqs.(B7)-(B9). 
}
\end{figure} 

\begin{figure}[t]
\includegraphics[width=1\linewidth]{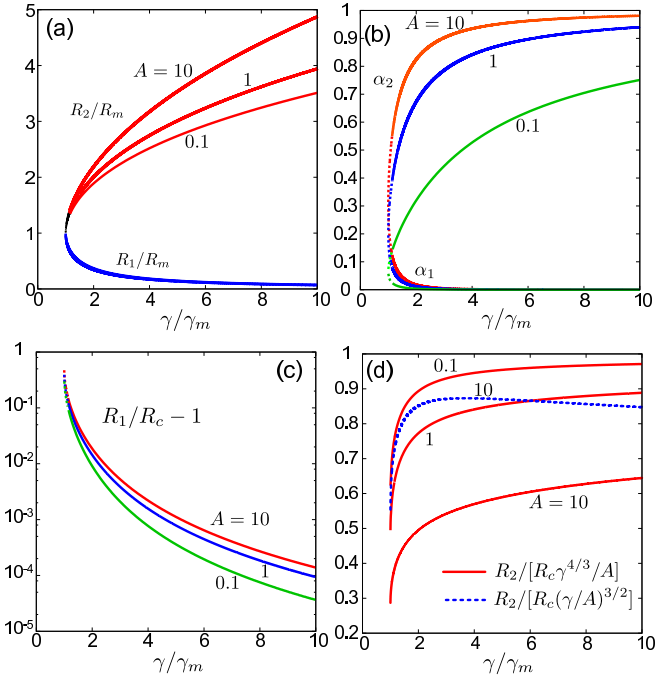}
\caption{(Color online) Results in the isobaric condition.  
(a) $R_1/R_{\rm m}= (s_1/s_{\rm m})^{1/3}$ 
and $R_2/R_{\rm m}= (s_2/s_{\rm m})^{1/3}$ vs $\gamma/\gamma_{\rm m}$ 
and (b) $\alpha_1= s_1/(1+ s_1)$ 
and $\alpha_2= s_2/(1+s_2)$ vs $\gamma/\gamma_{\rm m}$, 
where  $A=0.1, 1$, and 10.   
(c) $R_1/R_c-1=s_1^{1/3}\gamma/A-1$ vs $\gamma/\gamma_{\rm m}$, which is small 
even for vs $\gamma=\gamma_{\rm tr}$ (see eq.(43)). 
(d)  $R_2/[R_c \gamma^{4/3}/A]$ 
for $A=0.1$, 1, and 10 (bold lines) and 
$R_2/[R_c (\gamma/A)^{3/2}]$ 
for $A=10$ (dotted line) as functions of  $\gamma/\gamma_{\rm m}$ 
(see eqs.(44)-(46)).  
 }
\end{figure}

For each $A$, the  right hand side of eq.(B6)  is 
  minimized at $s=s_{\rm m}(A)$ as a function of $s$, 
 where $s_{\rm m}$ satisfies  
\be 
A= 3s_{\rm m}^{4/3}/(1-2s_{\rm m}). 
\en 
The  minimum of  eq.(B6)  at  $s=s_{\rm m}$ 
is written as 
\be 
\gamma_{\rm m}= s_{\rm m}(4+ s_{\rm m})/(1-2 s_{\rm m}). 
\en  
As can be seen in Fig.9(a), if $\gamma>\gamma_{\rm m}$, 
eq.(B6) has two solutions 
$s_{1}(\gamma, A)$ and $s_{2}(\gamma, A)$ with $s_1\le s_2$, 
where ${\cal F}(s,\gamma,A)$ exhibits 
   a local maximum ${\cal F}_{\rm max}(\gamma,A)$ at $s=s_{1}$ and a local 
minimum ${\cal F}_{\rm min}(\gamma,A)$  at $s=s_{2}$. Further 
 increasing $\gamma$ above $\gamma_{\rm m}$,  
the local  minimum ${\cal F}_{\rm min}$ decreases 
and becomes negative  
for  $\gamma> \gamma_{\rm tr}(A)$, 
 where $\gamma_{\rm tr} $ and the corresponding 
$s$, written as $s_{\rm tr}(A)$, are calculated from  
\bea
&&\hspace{-1cm}
A/(A+s_{\rm tr}^{1/3}) = 2(1+s_{\rm tr}^{-1}) \ln (1+s_{\rm tr})-2,\\
&&\gamma_{\rm tr} 
= s_{\rm tr} + A (1+s_{\rm tr} )/s_{\rm tr}^{1/3}. 
\ena 
See Fig.4(a) for   $\gamma_{\rm m}$ and $\gamma_{\rm tr}/\gamma_{\rm m}( 
\cong 1.14)$ vs $A$. 

From eqs.(B7)-(B10), we seek the asymptotic behaviors 
for small and large $A$. For $A\ll 1$  we find   
\bea 
&&\hspace{-1cm}
s_{\rm m} \cong (A/3)^{3/4}, \quad~ s_{\rm tr} \cong A^{3/4}\nonumber\\
&&\hspace{-1cm}
\gamma_{\rm m} \cong 4(A/3)^{3/4}, \quad \gamma_{\rm tr} \cong 2A^{3/4}.
\ena 
On the other hand, for $A\gg 1$,  we have 
\bea 
&&\hspace{-1cm}
s_{m} \cong 1/2, 
\quad\quad~~~~~~  s_{\rm tr} \cong 1.401  \nonumber\\ 
&&\hspace{-1cm}
\gamma_{m} \cong (3/2^{2/3})A, 
\quad\gamma_{\rm tr} \cong 2.145 A, 
\ena 
where $s_{\rm tr}$ and $\gamma_{\rm tr}$ are calculated numerically. 
If $\gamma\gg \gamma_{\rm m}$, we obtain 
eqs.(43)-(46).  

For $\gamma>\gamma_{\rm m}$,  we consider the radii 
$R_{\rm m}$, $R_1$, and $R_2$ 
corresponding to $s_{\rm m}, s_1$,  and $s_2$. 
From eq.(B2)  we have 
\be 
\frac{R_{\rm m}}{ R_0}=  s_{\rm m}^{1/3}, \quad 
\frac{R_1}{ R_0}=  s_1^{1/3}, \quad 
\frac{R_2}{ R_0}=  s_2^{1/3}, 
\en 
where $R_0=  2\sigma /[({\bar p}- p_{\rm cx})A]= R_c \gamma/A$. 
In Fig.10, we plot 
$R_1/R_{\rm m}= (s_1/s_{\rm m})^{1/3}$ and 
$R_2/R_{\rm m}=(s_2/s_{\rm m})^{1/3}$ vs $\gamma/\gamma_{\rm m}$ in (a) and  
  $\alpha_1= s_1/(1+ s_1)$ 
and $\alpha_2= s_2/(1+s_2)$ vs $\gamma/\gamma_{\rm m}$ in (b), 
where the latter are the solute fractions in bubbles 
at $R=R_1$ and $R_2$. Furthermore, in (c), $R_1/R_c-1$ is shown to be small 
for $\gamma$ slightly larger $\gamma_{\rm m}$  in accord with eq.(42). 
In (d), we divide $R_2/R_c$  by its 
asymptotic forms for $\gamma\gg \gamma_{\rm m}$ 
to confirm  eqs.(44)-(46).

\vspace{2mm}
{\it  2. Fixed volume }\\
We next scale the bubble free energy $\Delta F$ 
in  the fixed-volume condition    in Appendix A. 
We assume  the pressure balance (A2) and use $\hat\omega$ in eq.(A4). 
Introducing  the scaled 
bubble free energy 
 $\cal F$ as in eq.(B3) (with replacement $\Delta G \to \Delta F$), 
we express it in terms of $s$ in eq.(B1),  
$v$ in eq.(B2), and  
\be 
\hat{v}=(2\sigma /R+ \phi/K_T)/({\bar p}-p_{\rm cx})= 
v+s/h,
\en 
where $h$ is defined in eq.(A6). 
Replacing  $v$  by $\hat v$ in the first two   terms in eq.(B4),  
we obtain   
\bea 
&&\hspace{-6mm}
{\cal F}
=  [\gamma+1-s(1+\hat{v}) ] \ln\bigg[1-s\frac{1+\hat{v}}{1+\gamma}\bigg ]
 \nonumber\\
&&\hspace{-6mm} +s(1+ \hat{v}) \ln \bigg[\frac{1+\hat{v}}{1+\gamma}\bigg]
+  \frac{3}{2}sv +s +  \frac{s^2}{2h},
\ena  
where the last term arises from the compression term in eq.(A4). 
As in   eq.(B5), the  derivative of 
$\cal F$ with respect to $s$ is calculated as 
\be 
\frac{\p}{\p s}{\cal F}= 
\bigg(1+ \frac{2v}{3}+ \frac{2s}{h}\bigg)
\ln \bigg[\frac{1+\hat{v}}{\gamma+1-s(1+\hat{v})}\bigg ].
\en 
The extremum  condition ${\p}{\cal F}/\p s=0$ yields 
\bea 
\gamma &=& s+ (1+s)\hat{v}\nonumber\\
&=& s+ A (1+s)s^{-1/3}+ (s+s^2)/h ,
\ena 

As in the fixed pressure case, ${\cal F}(s,\gamma, A,h)$ exhibits 
a local maximum at $s=s_1(\gamma,A,h)$ and a local minimum 
at $s=s_2(\gamma,A,h)$ for $\gamma>\gamma_{\rm m}(A,h)$ 
and its local minimum becomes negative for $\gamma>\gamma_{\rm tr}(A,h)$.  
Note that the   right hand side of eq.(B17) 
is minimized at $s=s_{\rm m}(A,h)$, where  $s_{\rm m}$ is determined by 
\be 
A=3s_{\rm m}^{4/3}[ 1+(1+2s_{\rm m})/h]/({1-2s_{\rm m}}).
\en 
The corresponding minimum of eq.(B17) is  written as  
\be 
\gamma_{\rm m}=[4+s_{\rm m}+ 4(1+s_{\rm m})^2/h] s_{\rm m}/(1-2s_{\rm m})
\en 
Here,  we assunme  $Ah\ll 1$, where $s_{\rm m} \sim (Ah)^{3/4}\ll 1$ 
from eq.(B18).   In fact, for O$_2$ in ambient water, 
we obtain $Ah=0.010 [n_{\rm b}G(\theta)]^{1/3}$ 
with $n_{\rm b}$ in units of $\mu$m$^{-3}$, 
which is independent of ${\bar p}-p_{\rm cx}$. 
Then, as in eq.(B11),  we find  the asymptotic behaviors,  
\bea 
&&\hspace{-1cm}
s_{\rm m} \cong (Ah/3)^{3/4}, \quad~ s_{\rm tr} \cong (Ah)^{3/4}\nonumber\\
&&\hspace{-1cm}
\gamma_{\rm m} \cong 4(A/3)^{3/4}/h^{1/4}, 
\quad \gamma_{\rm tr} \cong 2A^{3/4}/h^{1/4},
\ena 
where $\gamma_{\rm tr}/\gamma_{\rm m}\cong 1.14$. 
As in eqs.(43) and (44), 
the radii $R_1$ and $R_2$ behave for $\gamma\gg \gamma_{\rm m}$ as 
\be 
R_1 \cong R_c,\quad 
R_2 \cong R_c  \gamma^{4/3}h^{1/3}/A\gg R_c.
\en

\end{document}